# A study of variability induced by events dependency in microelectronic production

(presented at the 7th IESM Conference, October 11 – 13, 2017, Saarbrücken, Germany)

Kean Dequeant, Pierre Lemaire, Marie-Laure Espinouse
Univ.Grenoble Alpes, CNRS, Grenoble INP, G-SCOP,
F-38000 Grenoble, France.
Email: kean.dequeant@grenoble-inp.fr,
pierre.lemaire@grenoble-inp.fr,
Marie-Laure.Espinouse@g-scop.grenoble-inp.fr

Philippe Vialletelle
STMicroelectronics,
F-38926, Crolles Cedex, FRANCE
Email: philippe.vialletelle@st.com

*Abstract*—Complex manufacturing systems are subject to high levels of variability that decrease productivity, increase cycle times and severely impact the systems tractability. As accurate modelling of the sources of variability is a cornerstone to intelligent decision making, we investigate the consequences of the assumption of independent and identically distributed variables that is often made when modelling sources of variability such as down-times, arrivals, or process-times. We first explain the experiment setting that allows, through simulations and statistical tests, to measure the variability potential stored in a specific sequence of data. We show from industrial data that dependent behaviors might actually be the rule with potentially considerable consequences in terms of cycle time. As complex industries require strong levers to allow their tractability, this work underlines the need for a richer and more accurate modelling of real systems.

## I. Accurate modelling of variability and the independence assumption

Industry 4.0 is said to be the next industrial revolution. The proper use of real-time information in complex manufacturing systems is expected to allow more customization of products in highly flexible production factories. Semiconductor High Mix Low Volume (HMLV) manufacturing facilities (called fabs) are one example of candidates for this transition towards "smart industries". However, because of the high levels of variability, the environment of a HMLV fab is highly stochastic and difficult to manage. The uncontrolled variability limits the predictability of the system and thus the ability to meet delivery requirements in terms of volumes, cycle times and due dates.

Typically, the HMLV STMicroelectronics Crolles 300 fab regularly experiences significant mix changes that result in unanticipated bottlenecks, leading to firefighting to meet commitment to customers. The overarching goal of our strategy is to improve the forecasting of future occurrences of bottlenecks and cycle time issues in order to anticipate them through allocation of the correct attention and resources. Our current finite capacity projection engine can effectively forecast bottlenecks, but it does not include reliable cycle time estimates. In order to enhance our projections, better forecast cycle time losses (queuing times), improve the tractability of our system and reduce our cycle times, we now need accurate dynamic cycle time predictions.

As increased cycle-time is the main reason workflow variability is studied (both by the scientific community and practitioners, see e.g. [1] and [2]), what follows concentrates on cycle times. Moreover, the "variability" we account for should be understood as the potential to create higher cycle times, even though "variability" may be understood in a broader meaning. This choice is made for the sake of clarity, but the methodology we propose and the discussion we lead can be applied to any other measurable indicator.

Sources of variability have been intensely investigated in both the literature and the industry, and tool down-times, arrivals variability as well as process-time variability are recognized as the major sources of variability in that sense that they create higher cycle times (see [3] for a review and discussion). As a consequence, these factors are widely integrated into queuing formulas and simulation models with the objective to better model the complex reality of manufacturing facilities. One commonly accepted assumption in the development of these models is that the variables (MTBF, MTTR, processing times, time between arrivals, etc.) are independent and identically distributed (i.i.d.) random variables. However, these assumptions might be the reason for models inaccuracies as [4] points out in a literature review on queuing theory. Several authors have studied the potential effects of dependencies, such as [5] who studied the potential effects of dependencies between arrivals and process-times or [6] who investigated dependent process times, [4] also gives further references for studies on dependencies effects. In a previous work [3], we pinpointed a few elements from industrial data that questioned the viability of this assumption in complex manufacturing systems.

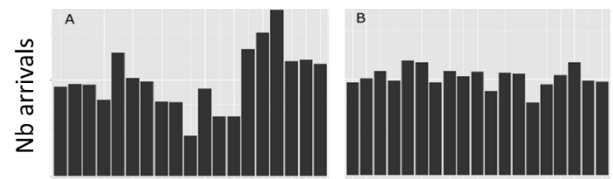

Figure 1: Number of arrivals per week from real data (A) and generated by removing dependencies (B)

For instance, Figure 1.A shows the number of arrivals per week at a toolset from STMicroelectronics Crolles300 fab, while Figure 1.B shows arrivals per week generated using the same inter-arrival rate as in Figure 1.A, but assuming independence of arrivals. The number of arrivals per week appears to be much more variable in reality than with the i.i.d. assumption.

From the (variability) modeling point-of-view, the i.i.d assumption means that all the information of many individual similar events (e.g., down-times) is contained within their distribution, and that the actual *sequence* of the events does not carry any additional information or, here, variability potential. That is why some events can be described by standard statistics, such as the mean and the variance, that completely define the statistical distribution. For simplicity of explanations, we will refer to sequences of i.i.d. random variables as i.i.d. sequences and the events they represent as i.i.d. events. The objective of this article is to investigate the impact that the i.i.d assumption may have on the quality of cycle time models and simulations by testing real world sequences of data. We therefore propose in the rest of our article an experiment framework for testing the variability potential of dependencies and show the results obtained on down-time distributions.

## II. An experiment framework for testing the variability potential of dependencies

### A. Objective: testing sources of variability for "dependency variability potential"

The objective we have is to investigate whether or not some information contained in the sequence of events might contribute to the variability. We therefore wish to investigate if there is any significant variability potential stored in the sequences of industrial data. Note that, from an operational perspective, if the events are not i.i.d but the result in terms of cycle time generated is similar, it can be considered that there is no useful information contained in the sequence and that the assumption is acceptable.

What we understand by the "variability potential" of a source of variability is its tendency to generate higher cycle times. For instance, queuing theory proposes using the coefficient of variation (the ratio between the standard deviation and the mean) for measuring the variability potential of arrivals (applied on inter-arrival times) and processing times (see e.g. Factory Physics [7]). To go further, we propose a framework to test if a specific *sequence* of data points carries a different variability potential than an i.i.d. sequence from the same data points. We will illustrate this approach by running a simulation model that integrates tool down-times, as they are the main source of variability in semiconductor manufacturing. Figure 1 shows a sequence of down-times and up-times of a specific tool from the Crolles 300 semiconductor fab recorded for a period of 1 year.

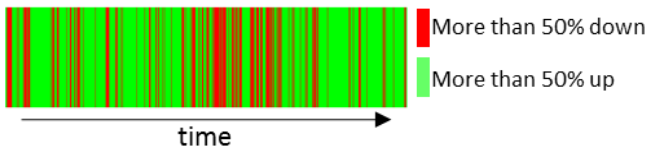

Figure 2: Historical down-times/up-times from a specific Crolles 300 tool over a year.

We call $S_0$ such a particular (real) sequence of events. Our fundamental question is whether the cycle time induced by the specific sequence $S_0$ is significantly different than that of an i.i.d. sequence from the same distribution, which we can create by applying a random permutation to the data points representing $S_0$. To answer this question, we propose an experiment framework that measures through simulations whether the effect of $S_0$ is statistically significantly different to that of i.i.d. sequences from the same distribution. The following subsections detail the different components of the proposed framework.

### B. Setting the simulation

The framework we propose is quite generic and could be applied to many situations: the simulation model could be any system, as long as one can control its inputs and measure its ouputs according to the hypothesis one wants to test.

In our case, we want to challenge the i.i.d. hypothesis, and test its effects on cycle-times. We therefore settle for a minimalistic simulation model, so as to disregard all potential sources of variability other than the independence (or not) of the down-times. We therefore consider the following system: agents queue in an infinite FIFO queue, waiting to be processed by a single tool with constant process times that treats each agent one after another with no inefficiency whatsoever except for the down-time events. These down-time events are set to follow specific sequences, being on the one hand the reference (real) sequence $S_0$, and on the other hand (as a basis for comparison) many other i.i.d. sequences. We ran all our simulations with Anylogic Professional in an agent based simulation exported as a standalone application: this allowed us to control the simulation model with R, and manage the inputs/ouputs of thousands of simulations in an automated manner.

### C. Measuring the effect of $S_0$

In our case, the effect of a sequence of down-times is measured as the average cycle-time of the agents in a simulation. This should be evaluated on an infinite horizon, that is until the measure is stable enough, but it is in practice impossible since $S_0$ is finite (as it is real, empirical data). As an alternative, to prevent any bias from a particular arrival sequence, we simply evaluate the long term average cycle-time by running many simulations on different scenarios, i.e. on different arrival sequences (following i.i.d. exponentially-distributed inter-arrival times). Thus, let $CT_{0,j}$ be the mean cycle time of agents from the simulation run on sequence $S_0$ of down-times and scenario $j$ of arrivals; the effect of $S_0$ is then $\overline{CT_0}$, the mean of mean cycle times from simulations that used $S_0$ as an input. The number of different scenarios $j$ is an important parameter and will be discussed in section II.F.

### D. Measuring the effect of i.i.d. sequences

To measure the effect of i.i.d. sequences, we generate many such sequences from the same distribution; those sequences ($S_{i,i\neq 0}$) are generated as random uniform permutations of $S_0$. Using the same procedure as for $S_0$, we compute for each sequence $S_{i,i\neq 0}$ its effect: $\overline{CT}_{i,i\neq 0}$. Note that we use the exact same scenarios for arrivals as for $S_0$, this extra requirement insuring that any significant difference between the population

$\overline{CT}_{i,i\neq 0}$ and the value $\overline{CT}_0$ comes strictly from the difference between $S_{i,i\neq 0}$ and $S_0$.

The effect of a particular i.i.d. sequence is of virtually no interest as we want to measure the effect of *any* i.i.d. sequence. This could be measured as the mean value of the $\overline{CT}_{i,i\neq 0}$, but to get a better control on the comparison we prefer measuring the effect as a 95% confidence interval $I_{95}$.

We compute $I_{95}$ as $\mu \mp 2\sigma$ where $\mu$ and $\sigma$ are respectively the mean and the standard-deviation of $\overline{CT}_{i,i\neq 0}$. Doing such calculation, we assume that the $\overline{CT}_{i,i\neq 0}$ follow a Gaussian distribution. We justify this statement both by the central limit theorem as each point of our sample population is a mean and by the fact that the underlying population $CT_{i\neq 0,j}$ most likely follows a Gaussian distribution (since the cycle time of each distribution comes from the accumulation of many independent random effects).

As for the computation of $\overline{CT}_0$, the right number of sequences and of scenarios must be carefully chosen so as to get good estimates within reasonable running times. This is also discussed in section II.F.

E. Comparing the effects

Once $\overline{CT}_0$ and $I_{95}$ have been computed accurately, they must/can be compared using a standard procedure for a statistical test: if $\overline{CT}_0$ falls within $I_{95}$, then it cannot be said that $S_0$ has an effect significantly different than that of an i.i.d. sequence; In the alternative, it can be assumed that $S_0$ has a different effect on cycle-time than an i.i.d. sequence. Figure 3 illustrates the three possible outcomes of the test. If $\overline{CT}_0$ is on the left side of $I_{95}$ (Figure 3 A), $S_0$ carries a negative variability potential; If $\overline{CT}_0$ is within $I_{95}$ (Figure 3.B), it cannot be said that $S_0$ carries any significant potential; If $\overline{CT}_0$ is on the right side of $I_{95}$ (Figure 3.C), $S_0$ carries a positive variability potential.

Note that setting a 95% confidence interval actually means, from a skeptic point-of-view, that there is a 5% chance for any sequence tested that the results will turn out to be positive by pure randomness. It is therefore essential to test different $S_0$ sequences, and to compare the number of positives with the probability of getting such results by "luck". For instance, if we test 19 sequences, the chance of getting more than half results positive by luck is less than $10^{-5}$.

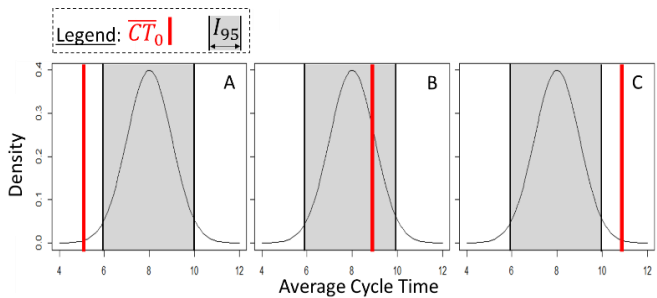

Figure 3: Three possible outcomes when comparing $\overline{CT}_0$ to the sample population $\overline{CT}_{i,i\neq 0}$

Note that a Student's t-test is not directly applicable here as it does not answer the question we are asking: should we perform a t-test between the populations $\overline{CT}_{0,j}$ and $\overline{CT}_{i\neq 0,j}$, we would be asking if the mean of both populations are identical or not. With a finite horizon of 1 year (resulting in a significant difference between each $S_i$), almost all $\overline{CT}_i$ are different than the mean of $\overline{CT}_{i\neq 0,j}$, thus a significant difference with this test would not bring any particular information on $S_0$.

As a complementary measure, we check, for each arrival scenario $j$, whether the mean of the population $\overline{CT}_{i\neq 0,j}$ is statistically different than $\overline{CT}_{0,j}$. For each scenario $j$, we can answer this question using a t-test and then count the number of times a scenario $j$ gave $\overline{CT}_{0,j}$ respectively statistically smaller, non-statistically different and statistically higher than the mean of the population $\overline{CT}_{i\neq 0,j}$. As the population $\overline{CT}_{i\neq 0,j}$ represents average cycle times from identical simulation sets and parameters, only with randomness coming from different draws of the parameters, we can assume that they follow a normal distribution and can apply the t-test. These numbers are referred to as the "t-test triplets" in the results section III and are computed as a secondary test, keeping the previously explained methodology as the main measure.

F. Handling uncertainties on $\overline{CT}_0$ and $I_{95}$

As $\overline{CT}_0$ and $I_{95}$ come from a finite population we created through simulation, the values we measure are actually estimates of the true values (that we would get if we had an infinite population) and therefore carry uncertainty. The actual value of $\overline{CT}_0$ is contained in a confidence interval around $\overline{CT}_0$, which we call $[\overline{CT}_0]$. The same applies to $I_{95}$: as we had to estimate $I_{95}$, we are not certain of the two limits of this interval, however $I_{95}$ is most likely contained within inner/outer limits which we call $I_{95}^-$ and $I_{95}^+$.

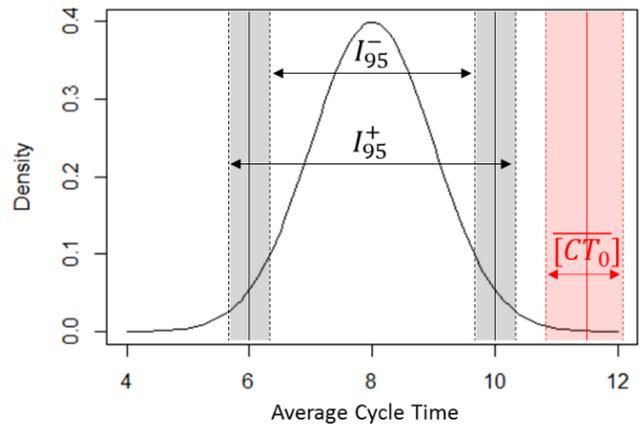

Figure 4: uncertainty areas around $\overline{CT}_0$ and $I_{95}$

These limits are represented on Figure 4 along with two greyed areas ($I_{95}^+ - I_{95}^-$) and the interval $[\overline{CT}_0]$. If $\overline{CT}_0$ is far away enough from $I_{95}$ as for $[\overline{CT}_0]$ and ($I_{95}^+ - I_{95}^-$) not to intersect, then whatever the value $\overline{CT}_0$ takes within $[\overline{CT}_0]$, and whatever the value $I_{95}$ takes within ($I_{95}^+ - I_{95}^-$), the conclusion is the same (as for Figure 3). The problem arises when the uncertainty areas of $[\overline{CT}_0]$ and $I_{95}$ overlap: depending on what true value $\overline{CT_0}$ and $I_{95}$ take inside their intervals, the answer can be different, and we can therefore not be sure of the answer. This is a "we do not know" situation as we cannot say with confidence that there is a difference, nor can we say with confidence that there is no difference.

In order to get definitive answers, these uncertainty areas around $\overline{CT_0}$ and $I_{95}$ need to be reduced to a point where they do not overlap anymore: to this extend, we need to increase the number of simulation runs so as to decrease either $[\overline{CT_0}]$ or the gap between $I_{95}^+$ and $I_{95}^-$. The appendix gives more details on the computation of these values and the way to minimize the number of simulations in order to get out efficiently of an overlap situation. In any case, there needs to be a minimum amount of simulations in order to start computing the uncertainty intervals. To test one $S_0$ sequence, we decided to run simulations on at least 20 i.i.d. sequences and 20 arrivals scenarios. Therefore, one simulation set is composed of at least 400 simulations.

G. Ending simulations

Each simulation needs to be run over the period of time defined by the sequence $S_0$ that is tested. However, at the end of this period of time, some agents may have accumulated in the simulation. Cutting the simulation abruptly can either lead to the loss of the cycle time of these agents or an underestimation of their cycle time depending on how the information is recorded in the simulation.

In order to prevent from such effects, we decided to stop the arrivals after the time duration defined by $S_0$, and continue the simulation until all agents have exited the simulation. We also chose to leave the tool up after the end of the given up/down sequence. This was first motivated by the fact that any other decision could introduce uncontrolled biases to the results. Secondly, by having a similar end to all simulations, we actually slightly underestimated the difference between the historical sequence and its i.i.d. counterparts: this only strengthens any significant results.

III. Results and implications

A. Down-times of C300 tools

Using the method described in section II, we tested 19 tools from the C300 fab. For each tool, we extracted the empirical sequence of up-times/down-times over a year. We then applied the procedure described in section II. One implicit parameter for all simulations is the utilization rate of the tool: in our simulation this utilization rate is fixed by adjusting the mean inter-arrival time of the scenarios. Since this utilization rate influences the cycle times, we set it as the same value across all simulations: 80%, as this is both a high and common, realistic level.

Table 1 summarizes the results of our experiments. The results are normalized so that the average cycle time for the sample i.i.d. sequences ($\mu$) is equal to 100. Therefore, the values of $\overline{CT_0}$ can be compared straightforwardly between them and relative to the central cycle time for i.i.d. sequences $\mu$. We also provide the confidence interval $[\overline{CT_0}]$ and the $I_{95}^*$ value, which correspond to $I_{95}^+$ if $\overline{CT_0}$ is outside $I_{95}^+$ or $I_{95}^-$ otherwise. The column "s" shows the result of our test: YES if there is a significant different, NO if there is no significant different, and NA if we were not able to get a definitive answer. We also add the "t-triplets" described in section II.E. Tools with same letters are from the same tool type.

| TOOL | $\overline{CT_0}$ | $\overline{CT_0} - \mu$ | $I_{95}^*$ | $[\overline{CT_0}]$ | s | t-test triplets |
|---|---|---|---|---|---|---|
| A1 | 78 | -22% | 81-119 | 76-80 | YES | (21-0-0) |
| A2 | 80 | -20% | 83-117 | 77-83 | YES | (28-0-0) |
| B3 | 175 | 75% | 45-155 | 165-185 | YES | (0-0-20) |
| C4 | 115 | 15% | 88-112 | 113-118 | YES | (3-8-89) |
| D5 | 668 | 568% | 52-148 | 645-691 | YES | (0-0-20) |
| E6 | 128 | 28% | 63-137 | 119-137 | NO | (0-0-21) |
| F7 | 182 | 82% | 60-140 | 176-189 | YES | (0-0-20) |
| F8 | 301 | 201% | 60-140 | 279-324 | YES | (0-0-20) |
| G9 | 131 | 31% | 75-125 | 126-136 | YES | (0-0-29) |
| H10 | 111 | 11% | 88-112 | 109-113 | NA | (10-11-79) |
| I11 | 144 | 44% | 71-129 | 136-153 | YES | (0-0-20) |
| J12 | 156 | 56% | 50-150 | 151-161 | YES | (0-0-20) |
| K13 | 208 | 108% | 78-122 | 191-225 | YES | (0-0-23) |
| K14 | 1385 | 1285% | 40-160 | 1348-1421 | YES | (0-0-20) |
| K15 | 261 | 161% | 79-121 | 246-276 | YES | (0-0-26) |
| K16 | 240 | 140% | 62-138 | 229-250 | YES | (0-0-20) |
| K17 | 420 | 320% | 66-134 | 405-436 | YES | (0-0-20) |
| L18 | 123 | 23% | 66-134 | 119-128 | NO | (0-0-20) |
| M19 | 104 | 4% | 92-108 | 100-107 | NO | (18-4-30) |

Table 1: Results of the experiment framework tested on 19 tools from STMicroelectronics

As Table 1 shows, out of the 19 historical sequences we tested, 3 did not show any significant difference in the variability potential they carried compared to their i.i.d. counterparts, 13 showed an increase, 2 showed a decrease, and 1 could not draw a definitive answer after 10000 simulation runs. The t-test triplets confirm the results and bring additional information on the impact of the arrival scenarios, but are not enough to assert a significant difference (e.g. see tools E6 and L18).

The differences between $\overline{CT_0}$ and $\mu$ in Table 1 first show that the empirical sequences of down-times we tested are statistically different than i.i.d sequences in terms of variability potential. Secondly, it should be noticed that the historical sequences added non negligible cycle time: the average increase in cycle time when the variability potential is positive is +174%, the median increase being +64%. This means that, for the majority of cases we tested, more than one third of the cycle time came from the actual sequence and not from the statistical distribution.

An interesting result is that the sequence of down-times can actually carry negative variability potential. The probability of false-positive is here ruled out by the fact that both tools that showed a sequence of down-times with a negative variability potential are from the same tool type. We explain this as being the result of "regular" down events amongst "irregular" down events. Indeed, these tools need regular changes of an important resource which create down-time events. However, even though the time between these special events are somewhat constant, there are many other down-time events that happen more or less randomly. Therefore the regularity of these events is lost in the statistics, and when running a simulation with i.i.d. down-time events, this regularity is broken and more variability occurs. The results from table 1 were obtained at a utilization rate of 80%. However, it is also interesting to run distinct experiments for the same sequences, but at different utilization rates. For each empirical sequence, we actually ran experiments on 9 different utilization rates ranging from 10% to 90% for a total number of 180 experiments, each requiring a minimum of 400 simulations. Figures 5 shows operational curves (representing cycle time versus utilization rate) for tools showing respectively an increase in variability potential, no significant change in the variability potential, and a decrease in the variability potential.

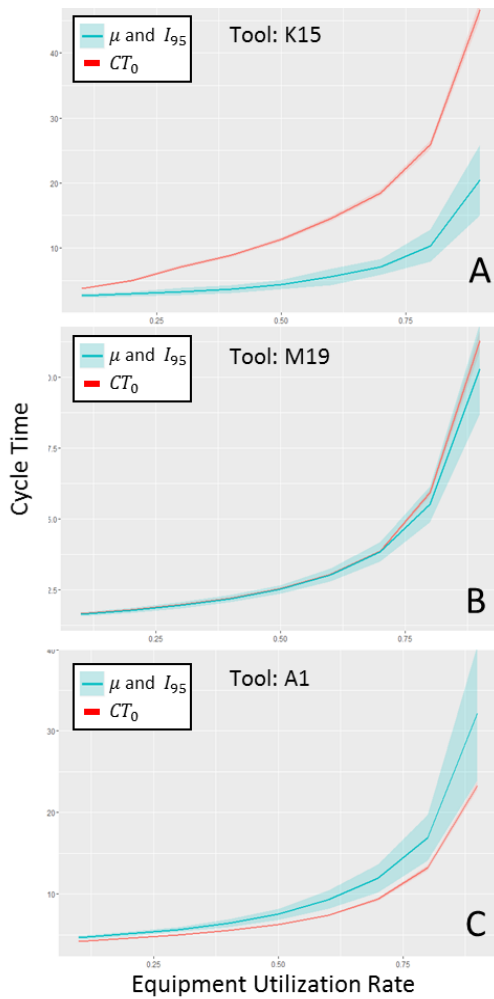

Figure 5: operational curve showing $\overline{CT_0}$ (statistically) higher (A), non-different (B), and lower (C) than $\mu$

The blue curves of Figures 5 are the average cycle times for the sample populations $\overline{CT}_{i, i \neq 0}$ along with its 95% confidence interval. They represent the outputs of the simulations run on the i.i.d down-times sequences. The red curves represent the values $\overline{CT_0}$, i.e. the average cycle times generated by the simulation runs that used the historical down-times sequences. These curves display the expected exponential behavior of cycle time when it comes to utilization rates. The cycle times added (or removed) by the dependencies seem to be proportional to the original cycle times and the utilization rate does not seem to impact whether $S_0$ is significantly different or not.

B. Results from an operational point-of-view

The results we show indicate that a large part of the cycle time generated by down-time events from historical data in the case of i.i.d. arrivals is unaccounted for in most queuing theories and simulations. However, the difference in cycle times between most current models (using queuing theory or simulation) and reality is not in the order of magnitude that we pointed out. How can it be that we see a significant difference that does not happen in reality?

We believe this difference comes from another dependency phenomenon that is still mostly unaccounted for: the dependency between the arrivals and the tool behaviors. Indeed, in the complex reality of manufacturing, decisions are made dynamically based on the flow of products. For instance, maintenances might be pushed back in the case of temporary over-saturations, more maintenance staff is also affected to tools that are forecasted to be highly used in the near future, etc.

IV. Perspectives and conclusion

A. Quantifying other sources of variability

The results we give in this article show strong evidence that tool down-times actually do not follow i.i.d. distributions and that this difference has a huge impact on the variability potential of down-times. As Fig. 1 shows, there is strong evidence that other sources of variability such as arrivals and processing-times also do not follow i.i.d. sequences. The next step is therefore to apply the same procedure on historical arrivals sequences as well as historical processing-times sequences.

B. Quantifying intra-sources relations

As we previously mentioned, there is strong evidence for a relation between the different sources of variability both from the shop-floor experience and from our quantification of the variability potential of down-time events. Further research should focus on quantifying this relation. Indeed, better modelling this relation could help both making models more accurate and reduce the overall variability fabs experience, as measuring is always the first step to reducing.

C. Conclusion

Accurate modelling of sources of variability (i.e. root causes for the generation of queuing time inefficiencies) is a key element of the manufacturing strategies put forward with industry 4.0 and High Mix Low Volume production. Previous works had questioned the viability of the assumption of independent and identically distributed random variables when it comes to the modelling of sources of variability such as tool down-times, arrivals, process times…

In this article, we proposed an experiment framework based on the repetition of many simulation runs that allows testing on historical sequences of data if this assumption has any implication in the modelling of variability. The experiment framework was built around the specific scenario of testing down-times, but can be applied to any sequence of data as long as a simulation is able to integrate the source of variability that is tested.

We then tested 19 different industrial down-time sequences of 1 year taken from tools from the Crolles 300 semiconductor fab. The results show that not only do the sequences of down-times carry significant variability potential, they actually contain an important part of the variability potential for most tools. We also showed that, quite interestingly, the sequence of down-times can actually carry negative variability potential and generate less cycle time than if they were sequences of independent and identically distributed variables.

Our results suggest that there is a strong relation between the arrivals and the behaviors of the tools in the reality of complex manufacturing. A next essential step is therefore to measure, quantify and understand the interactions that happen in the real world of complex manufacturing in order to better model these interactions and give industrials the levers they need to follow their transition to industry 4.0.

## Appendix

We mentioned in section II.G that the uncertainties on $[\overline{CT_0}]$ and $I_{95}$ might be too high to draw a definitive conclusion. As the uncertainties on $[\overline{CT_0}]$ and $I_{95}$ come from the estimators, they reduce with the number of simulation runs. Both uncertainty areas can be estimated straightforwardly: the standard deviation of the estimator for the mean $\mu$ and the standard deviation $\sigma$ are respectively $\frac{\sigma}{\sqrt{n}}$ and $\frac{\sigma}{\sqrt{2(n-1)}}$ where $n$ is the sample size [8]. Once again, we use these formulas as we make the assumption that the underlying populations follow normal distributions. Thus, we can define $[\overline{CT_0}]$ as $\overline{CT_0} \mp \frac{2\sigma_0}{\sqrt{m}}$, $I_{95}^-$ as $\mu \mp \left(2\sigma - \frac{\sigma}{\sqrt{n}} - \frac{2\sigma}{\sqrt{2(n-1)}}\right)$, and $I_{95}^+$ as $\mu \mp \left(2\sigma + \frac{\sigma}{\sqrt{n}} + \frac{2\sigma}{\sqrt{2(n-1)}}\right)$ where $\sigma_0$ and $\sigma$ are respectively the standard deviation of $\overline{CT_{0,j}}$ and $\overline{CT_{i,i\neq 0}}$; $n$ and $m$ are respectively the number of different i.i.d. sequences and of arrival scenarios; And $\mu$ is the mean of all $\overline{CT_{i,i\neq 0}}$. To derive the formulas on $I_{95}^-$ and $I_{95}^+$, we start with the estimation of $I_{95}(\mu \mp 2\sigma)$, and we subtract ($I_{95}^-$) or add ($I_{95}^+$) a standard-error for $\mu$ $\left(\frac{\sigma}{\sqrt{n}}\right)$ and two standard-errors for $\sigma$ $\left(\frac{\sigma}{\sqrt{2(n-1)}}\right)$.

After the initial 400 simulations, the condition to either increase the number of sequences $S_i$ or the number of arrivals scenarios is actually the same: an overlap between the red area and either of the grey areas of Figure 5. The decision of which to do first should be taken in order to minimize the computation time required. Drawing another arrival sequence will reduce $[\overline{CT_0}]$ whereas drawing another $S_i$ sequence will reduce the gap between $I_{95}^-$ and $I_{95}^+$. We therefore need to compute the decrease in the uncertainty area per simulation for either case of making a new draw of arrivals or of down-times, knowing that any new draw of arrivals needs to be run on all existing $S_i$ sequences and vice-versa.